\documentclass[conference]{IEEEtran}
\ifCLASSINFOpdf
\else
\fi
\usepackage{mathrsfs}
\usepackage{amsthm}
\usepackage{makeidx}
\usepackage{examples}
\usepackage[pdftex]{graphicx}  
\usepackage{amssymb}
\usepackage{algorithm}
\usepackage{epstopdf}
\usepackage{color}
\usepackage{url}
\usepackage{algorithm}
\usepackage{algorithmic}
\usepackage{amsmath}
\usepackage{hyperref}
\usepackage[utf8]{inputenc}
\def \x {{\bf x}}

\begin{document}
\title{3DNA Printer: A Tool for Automated DNA Origami}

\author{
\IEEEauthorblockN{Amay Agrawal, Birva Patel, Dixita Limbachiya and Manish K. Gupta\\}
\IEEEauthorblockA{Laboratory of Natural Information Processing \\ Dhirubhai Ambani Institute of Information and Communication Technology
Gandhinagar, Gujarat, 382007 India\\
Email: amayagrawal22@gmail.com, birva1@gmail.com, dlimbachya@acm.org, mankg@computer.org}
%\and
%\IEEEauthorblockN{Manish K. Gupta}
%\IEEEauthorblockA{Laboratory of Natural Information Processing\\
%Dhirubhai Ambani Institute of Information\\and Communication Technology\\
%Gandhinagar, Gujarat, 382007 India\\
%Email: mankg@computer.org}
}


\maketitle

\begin{abstract}
In the last two decades, DNA self-assembly has grown into a major area of research attracting people from diverse background. It has numerous potential applications such as targeted drug delivery, artificial photosynthesis etc. In the last decade, another area received wide attention known as DNA origami, where using M13 virus and carefully designed staple strands one can fold the DNA into desired 2-D and 3-D shapes. In 2016, a group of researchers at MIT have developed an automated DNA nanostructures strategy and an open source  software 'daedalus' based on MATLAB for developing the nanostructures. In this work, we present a truly open source software '3dnaprinter' based on Java (without MATLAB) that can do the same work. 
%We have optimized the algorithms used in previous work for efficiency. 
\end{abstract}


% \fi
\IEEEpeerreviewmaketitle

%%%%%%%%%%%%%%%%%%%%%%%%%%%%
\section{Introduction} 
%\textbf{Need to write}
Apart from being the blueprint of the life, DNA has been used for different applications \cite{wilner2012functionalized}. One of the application is to built nanostructures from DNA. DNA consists of four main bases namely A (Adenine), T (Tyhmine), G (Guanine) and C (Cytosine), in which G and C while A and T are complementary bases and they form double helix when they come together via hybridization. In 1980s, this intrinsic property was utilized by Ned Seeman and his collaborators to construct many interesting DNA structures at the nanoscale by trial and error such as DNA cube, DNA octahedron, DNA double crossover (DX) molecules etc \cite{seeman1982nucleic}, which gave birth to a field called structural DNA nanotechnology. In 1995 Erik Winfree explained that the self-assembly of DNA is Turing-universal \cite{winfree1998algorithmic}. That means, in principle, rather than trial and error one can systemically design any arbitrary shape with DNA. In a seminal paper in 2006, Rothemund introduced a method called DNA origami \cite{rothemund2006folding}, which is essentially the art of folding using DNA. By using all these self-assembly techniques one can design various interesting shapes such as smiley face, star, map of countries etc. In DNA Origami method, one can take a long single "scaffold" strand, that on hybridization at specific positions using "staple" strands results into the desired nanostructure.Since then, DNA nanostructures were built using distinct approaches as DNA tiles \cite{wei2012complex}, DNA bricks\cite{ke2012three}. The systematic approach was proposed by Peng Yin et al. to construct any arbitrary 2-D shape by DNA using DNA Tiles as the basic building block. Based on this concept, a software DNA Pen was developed \cite{goyal2013dna}. Scaling up this 2D model for 3D shapes, Peng Yin and his colleagues proposed the idea of DNA bricks which works like lego modeling to build the 3D structures. A software 3DNA is developed on this concept which allows designing of the 3D shapes by sculpting of DNA nanostructures \cite{gupta20143dna}. DNA nanostrucutres can be used in applications such as biosensing \cite{tai2008high}, triggered drug delivery \cite{linko2015dna}, enzyme cascades \cite{wilner2009enzyme}, bio-molecular analysis platforms \cite{cheng2006nanotechnologies} etc. 
%These approaches can be categorized in two classes top-down and the bottom-up methods. Aforementioned work in DNA origami is based on the bottom-up approach. Software like DNA Pen, 3DNA have used the bottom-up approach to model the DNA structures.

In this work, we introduced the software called 3DNAprinter to design the 3D structures using the systematic algorithmic top-down approach. This software is motivated from the work of Veneziano et al \cite{veneziano2016designer}. A software called DAEDALUS has been developed on this idea \cite{veneziano2016designer}. The main difference between DAEDALUS and 3DNAprinter software, is that our software is an open source implemented in java while DAEDALUS is an open source implemented in MATLAB,  which is not an open source even though it is free to use. We have slightly modified the algorithm used in DAEDALUS for efficiency. This software basically takes in the .obj file of any 3d shape (current version handles platonic solids) and converts it automatically into corresponding DNA sequences using the structured algorithmic self assembly. The 3 main steps employed in formation of structure are spanning tree generation, Eulerian circuit formation and staples addition. 

The paper is organized as follows. \textit{Section 2} contains an algorithm used to build the structure. \textit{Section 3} discuss the GUI of our software. \textit{Section 4} contains functionality and work-flow of our software. \textit{Section 5} contains few examples that have been build using our software. \textit{Section 6} provides the link from where our software is available. \textit{Section 7} concludes the paper with some general remarks.

%%%%%%%%%%%%%%%%%%%%%%%%%%%%%%%%%%%%%%%%%%%%%%%%%%%%%%%%%%%%%%%%%%%%%%%%%%%%%%%%%%%%%%%%%%%%%%%%%%%%%%%%%%5
\section{Algorithm}
\begin{figure*}[ht]
\centering
\includegraphics[scale=0.5]{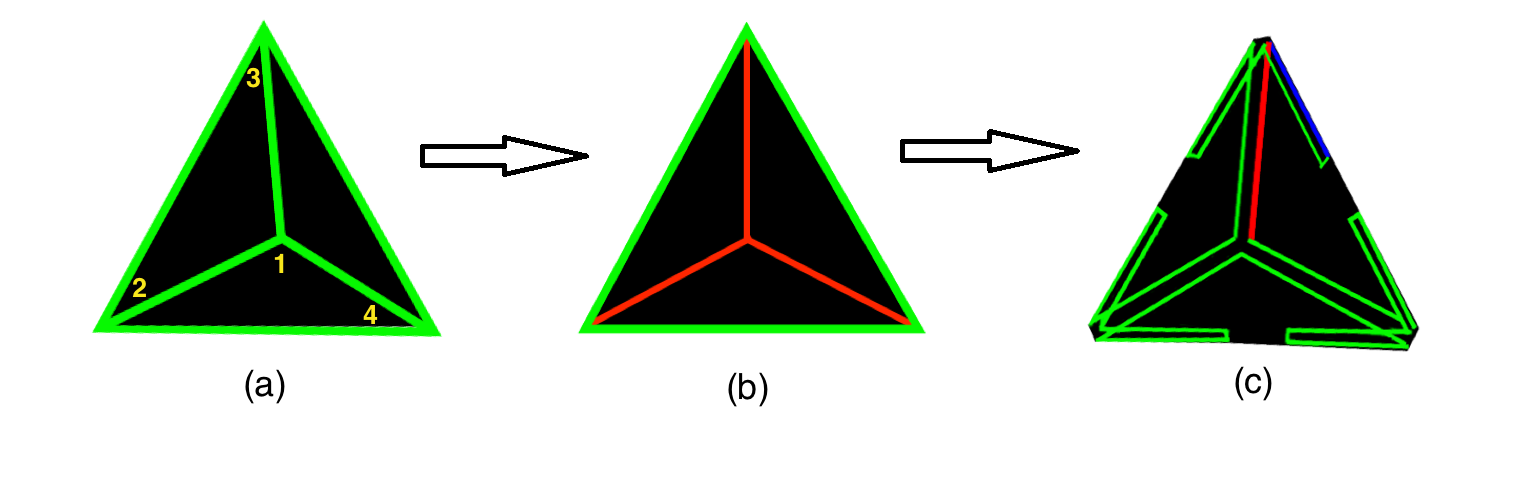}
\caption{(a) Shows the polyhedra mesh generated after importing the file.
(b) Shows the generation of the spanning tree from Polyhedra Mesh. Red edges are part of spanning tree. (c) Shows the Eulerian Circuit formation of Polyhedra Mesh. Red egde indicates from where you can start scaffold routing while blue indicates the second edge in routing. }
\label{gui}
\end{figure*}

In this section, we recall the algorithm used in \cite{veneziano2016designer} for constructing 3D DNA origami. The main purpose of this algorithm is to model and synthesize a DNA scaffold and its staples for any given 3-D object using the top-down approach. For this purpose, one can visualize any 3-D object into its corresponding 3-D graph. The algorithm firstly generates a DNA scaffolding route which is followed by generation of staples which pull the structure together to give the required 3-D shape. The steps of the algorithm are schematically represented in Figure \ref{gui}

\subsection{Generating the spanning tree}

For the algorithm to successfully route the 3-D structure, the basic requirement is that the Eulerian Circuit exists. An Eulerian Circuit is guaranteed if the degrees of the vertices are even and for that purpose we have decided to use 2 duplexes per edge. 
There are a number of routing solutions which deliver an Eulerian Circuit but not all of them will result in an effective scaffold routing. Thus to produce an effective scaffold route we have employed the Prims algorithm for the minimum spanning tree.  As the 3-D objects have equally weighted edges, any of the minimum spanning tree algorithms like Prim's, Kruskal's, DFS, BFS etc. could be used. But as we know that branching tree self-assemble more reliably than linear trees \cite{pandey2011algorithmic}, hence we have used Prim's algorithm which gives breadth first search spanning tree with most branches.
We know that the crossovers can be used to strengthen the 3-D structure \cite{arias2012entropy}. Crossover can occur in staple strands or in scaffold strands. According to our algorithm, each edge can correspond to either zero or one scaffold crossover. The edges that are part of the spanning tree are assigned a zero scaffold crossover and the remaining edges are assigned one scaffold crossover. This can be seen in the Figure \ref{gui}.b

\subsection{Adding pseudo-nodes and scaffold routing}

On formation of spanning tree, the graph is converted to an Eulerian circuit. This is done  by splitting the non-spanning tree edges into two halves by adding a pair of pseudo-nodes and keeping all the spanning tree edges intact. 
Depending on the degree of the vertex, a set of pseudo-vertices are added to replace that vertex node.In consequence to this, each pseudo-vertex of degree N has N edges emerging from this and N faces between them. Each pseudo-node is placed such that it joins two bordering edges of a face and disconnects that face from other edges. This forms the Eulerian Circuit of the 3-D object through which the scaffold will route.

The circuit can be routed from two different directions, either in clockwise or anticlockwise direction. But for the stability purposes \cite{he2010chirality}, we will route in the anticlockwise direction. Thus now we have a directed Eulerian graph. This can be seen in Figure \ref{gui}.c

\subsection{Sequence generation and adding staples}

From the Eulerian circuit constructed above, we can now route the DNA strand across the scaffold. As edge should be a multiple of $10.5$ bps, we have employed to use $52$ bps for the spanning tree edges while $26$ for the non-spanning tree cut edges. 

Our algorithm used 3 different kinds of staples: vertex staples, zero scaffold crossover edge staples and one scaffold crossover edge staples. These staples can be easily created by taking the complimentary base pairs of the scaffolded DNA strand. We have created these staples using two stack algorithm. 

Vertex staples can be generated by traversing the Eulerian circuit using two stack algorithm. For each edge that is a part of vertex staple, we use $10$ bps which can be from either $3'$ or $5'$ end that is complementary to the corresponding scaffold base pair. 

For edge staples with one scaffold crossover, two $31$ bps - $32$ bps staples are placed across the crossover with $15$ bps - $16$ bps on either side of crossover to provide strong binding. While the remaining $20$ bps - $21$ bps corresponding to that edge has been already used in forming the vertex staples.

In edges with zero scaffold crossover, it has $42$-nts staple with a staple crossover at $21$ bps. The remaining $20$ bps - $22$ bps are employed to make another C-shaped staple.

\begin{figure}[ht]
\centering
\includegraphics[scale=0.8]{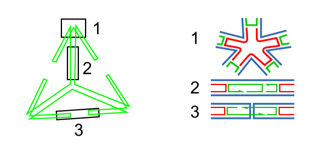}
\caption{1) Vertex Staple for tetrahedron. 2) Edge staple for tetrahedron. 3) Cut-edge staple for tetrahedron}
\label{staple}
\end{figure}

%%%%%%%%%%%%%%%%%%%%%%%%%%%%%%%%%%%%%%%%%%%%%%%%%%%%%%%%%%%%%%%%%%%%%%%%%%%%%%%%%%
\section{Graphical User Interface}
\begin{figure}[ht]
\includegraphics[width=8cm]{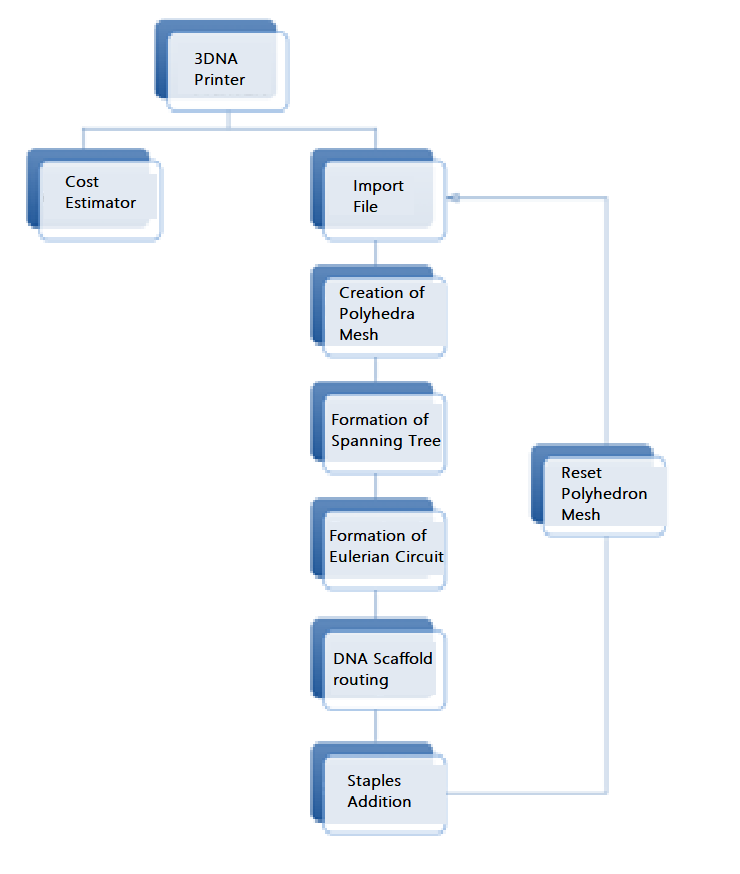}
\centering
\caption{Overview of 3DNAprinter GUI}
\label{flowdiagram}
\end{figure}

The GUI of the 3DNAprinter has been developed for the users to easily upload any 3D .obj file of any shape and get the corresponding DNA sequences of the shape to construct it at a nanoscale. Schematic representation of GUI is described in Figure \ref{flowdiagram}. 

\begin{figure}[h]
\centering
\includegraphics{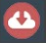}
\caption{Import button for importing .obj file}
\label{import}
\end{figure}

\subsection{Importing a file and Creating Polyhedra mesh}

As shown in the Figure \ref{import} one can find this button on the top left corner where one can import any 3D .obj file and the corresponding 3D object can be obtained as shown in Figure \ref{gui}.a

\subsection{Spanning Tree Generation of Polyhedra mesh}
\begin{figure}[ht]
\centering
\includegraphics{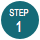}
\caption{Used for the spanning tree generation}
\label{step1}
\end{figure}

As shown in the Figure \ref{step1}  one can find this button on the top right corner where one can just click the button and the spanning tree corresponding to the 3D structure can be obtained as shown in Figure \ref{gui}.b

\subsection{Forming the Eulerian circuit for the scaffold path}
\begin{figure}[ht]
\centering
\includegraphics{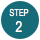}
\caption{Used for Eulerian Circuit formation}
\label{eulerian}
\end{figure}

As shown in the Figure \ref{eulerian} one can find this button on the top right corner where one can just click the button and see the Eulerian circuit which will then be used to route the scaffold along it. It is shown in Figure \ref{gui}.c

\subsection{Generating the DNA sequences corresponding to the structure}
\begin{figure}[ht]
\centering
\includegraphics{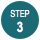}
\caption{Used for scaffold routing and generation of staple sequences and overall final sequences}
\label{step3}
\end{figure}

As shown in the Figure \ref{step3} one can find this button on the top right corner where one can just click the button and the DNA sequences corresponding to the imported 3D structure will be saved in excel file named $\mbox{Output}\_\mbox{Sequences.xls}$ which can be found in the sequences folder.

%%%%%%%%%%%%%%%%%%%%%%%%%%%%%%%%%%%%%%%%%%%%%%%%%%%%%%%%%%%%%%%%%%%%%%%%%%%%%%%%%%%%%%%%%%%%%%%%%%%%%
\section{Functionality and workflow}
The main aim of this subsection is to provide an overview of the working and functionality of the software. The key features of the software are its importing of the 3-D object files and output modules which include saving sequences corresponding to the 3-D object into the files.

%\textit{\textbf{a} Creating a polyhedra mesh}
\subsection{Creating a polyhedra mesh}
A new 3-D object file can be easily imported from the import file option. This 3-D object file (.obj extension) can be easily created in Auto-Cad, Maya etc. The user can then have a 3-D view of the corresponding object which is formulated by java3d libraries. The constructed polyhedron mesh can be seen in the figure \ref{gui}.a

\subsection{Spanning tree generation}

In this step, users can view the spanning tree of the corresponding 3-D object, which is made using prims algorithm. The edges that are a part of the spanning tree has been shown in red color which can be seen in figure \ref{gui}.b

\subsection{Scaffold routing path generation}

From the spanning tree, an Eulerian path is formulated through which the DNA strand will route the 3-D structure. This route is made by splitting the vertices into the pseudo-vertices corresponding to its degree and cutting the non-spanning tree edges. The figure \ref{gui}.c shows the entire Eulerian scaffold routing path where the starting edge of the Euler circuit is shown in red and the next edge in blue.

\subsection{Generating output sequences}

The set of sequences generated by the software are stored in a Excel sheet (.xls extension). The file contains the vertex-to-vertex DNA scaffold sequences, vertex staple sequences, edge staple sequences and cut-edge staple sequences corresponding to the 3-D object.

%\subsection{Checking thermodynamic stability}

%In this section we can check the thermodynamic stability of the 3D structure generated by our software corresponding to the given scaffold. First we calculate the entropy of the scaffold and the staple strands generated by us using the formula (G = H - T*S) and with the help of the table ----.From this we can conclude whether the scaffold taken and staples generated are individually stable or not. This stability is checked using the nearest neighbour method.If they all are found to be stable, then we again calculate entropy according to the paper [58] which is done by hybridization between the staple and scaffold strands and can find whether they are stable or not. If they are not stable we can change the scaffold and again repeat the whole process of sequence generation and then check its stability again. We have used M13 virus of 7429 bp as a default scaffold here. 

\subsection{Reset module}

The user can reset the canvas and import a new 3-D object file corresponding to a new structure.

\subsection{Estimator module}

This module can be used to calculate the amount based on the number of bps.

%%%%%%%%%%%%%%%%%%%%%%%%%%%%%%%%%%%%%%%%%%%%%%%%%%%%%%%%%%%%%%%%%%%%%%%%%%%%%%%%%%%%%%%%%%%%%%%%%%%%5
 
 \section{Examples}
 
Using 3DNA Printer, one can model various 3-D structures of varying degrees like a tetrahedron, octahedron, icosahedron etc. By default, we are using M13 virus of length $7429$ bps to route the scaffold although users can select the DNA strand of their choice. The details of constructing two 3-D structures namely tetrahedron and octahedron using our software is given as follows:

\begin{figure}[ht]
\centering
\includegraphics[width=8cm]{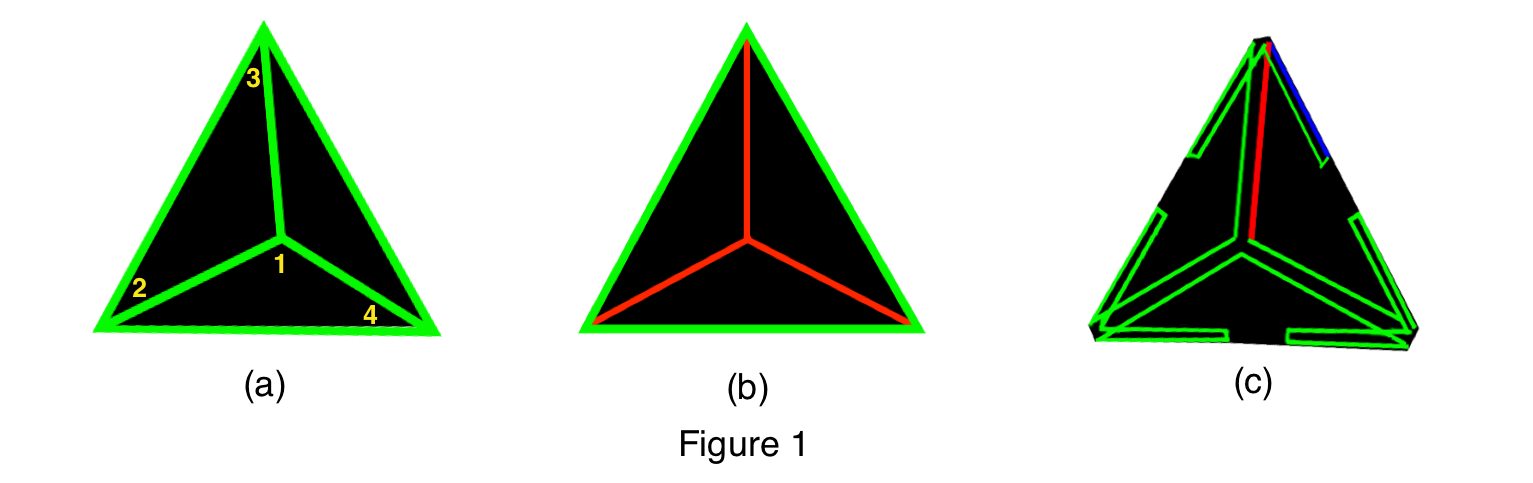}
\caption{Example of formation of scaffold and staple strands of Tetrahedron using our software}
\label{tetrahedron}
\end{figure}

\subsection{Tetrahedron} The pictorial representation of the tetrahedron can be seen in figure \ref{tetrahedron}, which is explained below: 
 
 \begin{itemize}
     \item A tetrahedron consists of 4 vertices which each vertex having a degree of 3, as shown in figure \ref{tetrahedron}.a
     
     \item The spanning tree of the tetrahedron is formulated using the prim's minimum spanning tree algorithm. As seen in the figure \ref{tetrahedron}.b, edges 1-2, 1-3, 1-4 are part of the minimum spanning tree which are represented in red color.
     
     \item The non-spanning tree edges i.e edges 2-3, 2-4, 3-4 are now split into two halves and whereas all spanning tree edges are kept intact. Also, each vertex is replaced by 3 pseudo-vertices to form a Eulerian circuit.
     
     \item Now after forming the Eulerian circuit, the default M13 is routed in the anticlockwise direction across the scaffold with each edge having 52 bp. As seen in figure \ref{tetrahedron}.c, starting edge is represented in red and the next edge in blue.
     
     \item Thus, we require a total of 4 vertex staples, 6 edge staples and 6 cut-edge staples for the formation of tetrahedron
     
     \item A total of $1026$ nucleotide bps is required for the formation of tetrahedron.
 \end{itemize}
 
 \begin{figure}[ht]
 \centering
\includegraphics[width=8cm]{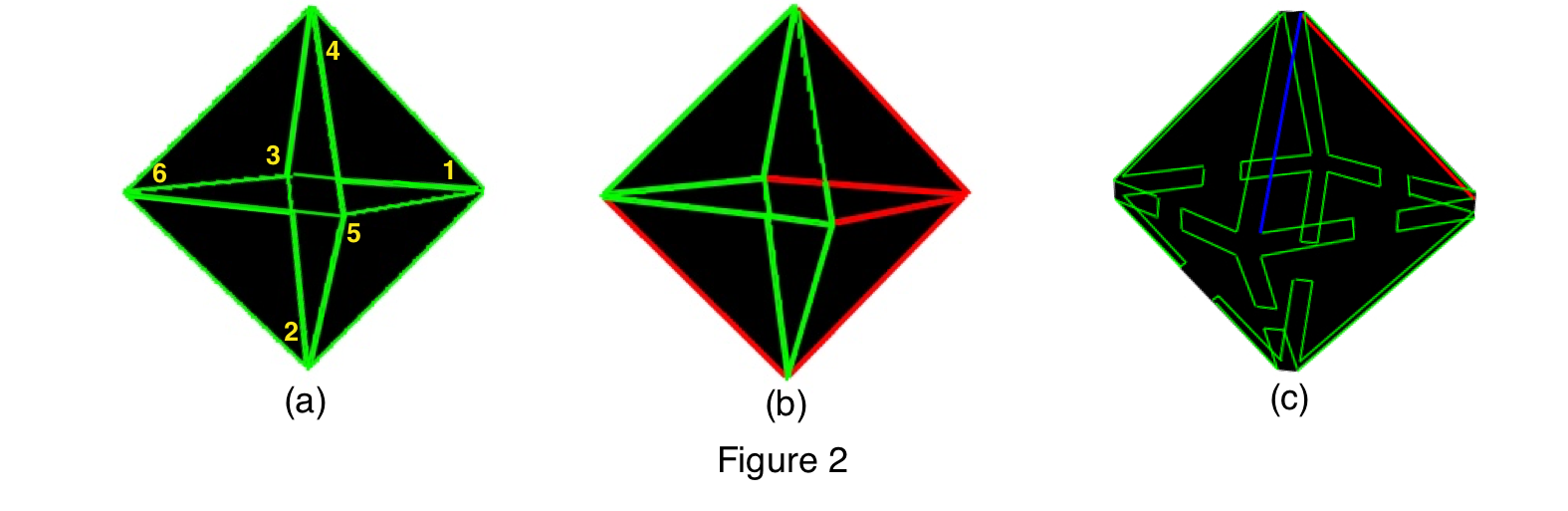}
\caption{Example of formation of scaffold and staple strands of Octahedron using our software}
\label{octahedron}
\end{figure}
 
 \subsection{Octahedron} 
    The pictorial representation of the tetrahedron can be seen in figure \ref{octahedron}, which is explained below: 
 
 \begin{itemize}
     \item An octahedron consists of 6 vertices which each vertex having a degree of 4, as shown in figure \ref{octahedron}.a.
     
     \item The spanning tree of the octahedron is formulated using the prim's minimum spanning tree algorithm. As seen in the figure \ref{octahedron}.b, edges 1-2, 1-3, 1-4, 1-5, 2-6 are part of the minimum spanning tree which are represented in red color.
     
     \item The non-spanning tree edges i.e edges 2-3, 2-5, 3-4, 4-5, 3-6, 4-6, 5-6 are now split into two halves and whereas all spanning tree edges are kept intact. Also, each vertex is replaced by 4 pseudo-vertices to form a Eulerian circuit.
     
     \item Now after forming the Eulerian circuit, the default M13 is routed in the anticlockwise direction across the scaffold with each edge having 52 bp. As seen in figure \ref{octahedron}.c, starting edge is represented in red and the next edge in blue.
     
     \item Thus, we require a total of 6 vertex staples, 10 edge staples and 14 cut-edge staples.
     
     \item A total of $2058$ nucleotide bps are required for the formation of tetrahedron.
 \end{itemize}
 
 %%%%%%%%%%%%%%%%%%%%%%%%%%%%%%%%%%%%%%%%%%%%%%%%%%%%%%%%%%%%%%%%%%%%%%%%%%%%%%%%%%%%%%%%%%%%%%%
 
 \section{Availibility}
 
 This software can easily be downloaded from the following URL http://www.guptalab.org/3dnaprinter/
 %%%%%%%%%%%%%%%%%%%%%%%%%%%%%%%%%%%%%%%%%%%%%%%%%%%%%%%%%%%%%%%%%%%%%%%%%%%%%%%%%%%%%%%%%%%%
 \section{Conclusion}
 This paper present a Java (without MATLAB) based open source software for developing 3D nanoscale structures automatically by importing a corresponding .obj file which can be obtained from any well known software like Autocad, Maya etc. for any desired 3D structure.Current version of the software designs only Platonic solids. In future, we will release a version which supports any 3D object.
 \bibliographystyle{IEEEtranS}
 \bibliography{3DNA}
\end{document}